\documentclass[a4paper,fleqn]{cas-sc}

\usepackage[authoryear,longnamesfirst]{natbib}
\usepackage{graphicx}
\usepackage{booktabs}
\usepackage{multirow}
\usepackage{longtable, array}

\def\tsc#1{\csdef{#1}{\textsc{\lowercase{#1}}\xspace}}
\tsc{WGM}
\tsc{QE}

\begin{document}
\let\WriteBookmarks\relax
\def\floatpagepagefraction{1}
\def\textpagefraction{.001}

\setlength{\tabcolsep}{5pt}

\shorttitle{The Effects of Generative AI Agents and Scaffolding on Enhancing Students' Comprehension of Visual Learning Analytics}

\shortauthors{Yan et al.}  

\title [mode = title]{The Effects of Generative AI Agents and Scaffolding on Enhancing Students' Comprehension of Visual Learning Analytics}  



%

\author[1]{Lixiang Yan}[orcid=0000-0003-3818-045X]
\cormark[1]
\author[1]{Roberto Martinez-Maldonado}[orcid=0000-0002-8375-1816]
\author[1]{Yueqiao Jin}[orcid=0009-0003-7309-4984]
\author[1,2]{Vanessa Echeverria}[orcid=0000-0002-2022-9588]
\author[1]{Mikaela Milesi}[orcid=0009-0002-0910-9822]
\author[1]{Jie Fan}[orcid=0009-0000-8585-2760]
\author[1]{Linxuan Zhao}[orcid=0000-0001-5564-0185]
\author[1]{Riordan Alfredo}[orcid=0000-0001-5440-6143]
\author[1]{Xinyu Li}
\author[1]{Dragan Gašević}

\affiliation[1]{organization={Monash University},
            addressline={25 Exhibition Walk}, 
            city={Clayton},
            postcode={3168}, 
            state={Victoria},
            country={Australia}}

\affiliation[2]{organization={Escuela Superior Politécnica del Litoral},
            city={Guayaquil},
            country={Ecuador}}            

\begin{abstract}
Visual learning analytics (VLA) is becoming increasingly adopted in educational technologies and learning analytics dashboards to convey critical insights to students and educators. Yet many students experienced difficulties in comprehending complex VLA due to their limited data visualisation literacy. While conventional scaffolding approaches like data storytelling have shown effectiveness in enhancing students' comprehension of VLA, these approaches remain difficult to scale and adapt to individual learning needs. Generative AI (GenAI) technologies, especially conversational agents, offer potential solutions by providing personalised and dynamic support to enhance students' comprehension of VLA. This study investigates the effectiveness of GenAI agents, particularly when integrated with scaffolding techniques, in improving students' comprehension of VLA. A randomised controlled trial was conducted with 117 higher education students to compare the effects of two types of GenAI agents: \textit{passive agents}, which respond to student queries, and \textit{proactive agents}, which utilise scaffolding questions, against standalone scaffolding in a VLA comprehension task. The results show that passive agents yield comparable improvements to standalone scaffolding both during and after the intervention. Notably, proactive GenAI agents significantly enhance students' VLA comprehension compared to both passive agents and standalone scaffolding, with these benefits persisting beyond the intervention. These findings suggest that integrating GenAI agents with scaffolding can have lasting positive effects on students' comprehension skills and support genuine learning.
\end{abstract}




\maketitle

\section{Introduction}

Technological advancements have enabled the capture and analysis of vast amounts of data, making visual analytics increasingly essential in education \citep{noroozi2019multimodal}. Visual learning analytics (VLA) are becoming more prevalent in learning analytics dashboards (LADs), classroom settings, and massive open online courses (MOOCs), serving as crucial tools for communicating insights to educators and students \citep{vieira2018visual}. For example, VLA is frequently used in learning analytics dashboards to display student collaboration metrics, heat maps of engagement levels, or progress tracking charts, which aim to provide students and educators with insights into learning behaviours and outcomes \citep{sahin2021visualizations}. The goal of VLA is to close the learning analytics loop by transforming raw data into actionable insights and feedback, thus guiding instructional strategies and enhancing educational experiences \citep{verbert2020learning}. 

Despite the potential of VLA, many students and educators struggle to interpret these complex visualisations due to a lack of data visualisation literacy \citep{donohoe2020data, pozdniakov2023teachers}. This challenge may intensify as VLA becomes more intricate with advancements in artificial intelligence (AI) and the integration of multimodal sensory data \cite{ochoa_multimodal_2022}. These developments not only increase the volume of educational data but also expand its dimensions and modalities, potentially resulting in more complex VLA that are even harder to comprehend \citep{echeverria2018driving, corrin2018evaluating}. Consequently, there is a growing need to support educational stakeholders' comprehension of VLA and ensure these analytics do not lead to cognitive overload, especially for students without much prior knowledge or exposure to complex VLA \citep{Ramaswami_2022, donohoe2020data}.

Supporting students' understanding of complex VLA could be achieved through the integration of explanatory and interactive features that scaffold their interaction with VLA \citep{echeverria2018driving, yan2024genai}. While conventional scaffolding approaches like data storytelling are effective in enhancing comprehension of VLA \citep{shao2024data}, these manual approaches remain difficult to scale and adapt to individual learning needs \citep{fernandez2024data, li2024we}. The advancement of generative artificial intelligence (GenAI), particularly with its multimodal capabilities, paves the way for GenAI-powered conversational agents. These agents can infuse interactivity into VLAs, transitioning from exploratory pattern identification to providing deeper, explanatory insights \citep{ooi2023potential, echeverria2018exploratory}. Evidence suggests that such GenAI agents offer real-time, personalised explanations, significantly easing the cognitive burden on students \citep{okonkwo2021chatbots, Kuhail_2022}. Additionally, the implementation of retrieval-augmented generation (RAG) techniques can elevate VLA from static information displays to dynamic, conversational systems. This transformation fosters a supportive learning environment where students can seek clarifications and obtain immediate, relevant responses \citep{verbert2020learning, yan2024genai}. Moreover, adopting an agentic system design can further evolve traditional AI systems (passive responsive) to proactive GenAI agents that actively navigate students through crucial components of VLA, thereby enhancing comprehension and facilitating improved learning outcomes \cite{gibbons2002scaffolding, kim2018effectiveness}. 

While GenAI agents offer promising opportunities to enhance students' comprehension of VLA \citep{yan2024vizchat, ma2023demonstration}, the empirical evidence supporting their effectiveness is still limited. To understand the impact of GenAI on students' comprehension of VLA, more rigorous studies are needed to distinguish between genuine learning gains and temporary performance boosts \citep{soderstrom2015learning, khosravi2023generative}. It is crucial to assess the tangible impact of GenAI agents on fostering a deep, lasting understanding of VLA, rather than merely facilitating a temporary grasp \citep{giannakos2024promise, cukurova2024interplay}. Although students might show improved comprehension due to the explanations and guidance provided by GenAI, these improvements may not persist without continued support and do not necessarily reflect an increase in actual competency and learning \citep{soderstrom2015learning, zimmerman2011self}. There is a risk that students could develop an "illusion of competence," where they believe they comprehend complex VLA through GenAI assistance without genuinely internalising the knowledge \citep{koriat2005illusions}. To address this issue, there is an urgent need for rigorous empirical research to evaluate the effects of GenAI agents on students' ability to comprehend complex VLA, particularly focusing on whether any improvements are sustained after the withdrawal of GenAI support.

This study aims to fill the gaps in the literature identified above by empirically evaluating the impact of \textit{three intervention conditions}: data storytelling, passive GenAI agents (those that can reactively respond to questions), and proactive GenAI agents (those that can proactively guide students to make sense of visual data by asking questions or providing suggestions) on students' effectiveness and efficiency in extracting insights from VLA. Specifically, we sought to quantitatively assess how each condition enhances accuracy and reduces the time required for insights extraction from visual data across \textit{three intervention phases}: pre-intervention, during the intervention, and post-intervention (after the intervention is removed). Including a post-intervention phase is crucial for understanding whether these supports function merely as temporary tools or also facilitate the learning of key comprehension skills of VLA during the process \citep{asamoah2022improving, bahtaji2020improving}. This motivates the first research question:

\begin{itemize}
\item \textbf{RQ1:} To what extent do students' \textit{effectiveness} and \textit{efficiency} in comprehending visual data \textbf{change} from pre- to post-intervention phases across the three intervention conditions?
\end{itemize}

We also aimed to compare the effectiveness and efficiency of the three interventions to identify the most effective method for sustaining comprehension improvements. This between-subject comparison is critical for understanding the relative efficacy of novel methods, such as GenAI agents, compared to state-of-the-art scaffolding approaches like data storytelling for enhancing VLA comprehension \citep{li2024we}. This leads to the second research question:

\begin{itemize}
\item \textbf{RQ2:} To what extent do students' \textit{effectiveness} and \textit{efficiency} in comprehending visual data \textbf{differ} between the three intervention conditions in pre-intervention, the intervention, and post-intervention phases?
\end{itemize}

Finally, considering the important role that visualisation literacy plays in VLA comprehension \citep{pozdniakov2023teachers, shao2024data}, we aimed to examine the effects of visualisation literacy on the three intervention conditions through the third research question:

\begin{itemize}
\item \textbf{RQ3:} To what extent do \textit{visualisation literacy} impact students' effectiveness and efficiency in comprehending visual data in each of the three intervention conditions?
\end{itemize}



The findings of this study contribute valuable insights into how GenAI agents, particularly proactive ones, can enhance students' comprehension of VLA by incorporating structured guidance, highlighting their potential to genuinely improve student competency beyond mere temporary performance enhancements.
\section{Background and Related Work}
\subsection{Visual Learning Analytics}

VLA is an evolving domain that intersects learning analytics and visualisation techniques to enhance the interpretation of complex educational data by stakeholders, including students, educators, and administrators \citep{vieira2018visual}. The primary aim of VLA is to decode vast amounts of educational data into intuitive visual forms that facilitate understanding and decision-making, thereby improving teaching practices and learning outcomes \citep{vieira2018visual, noroozi2019multimodal}. The significance of VLA lies in its ability to uncover insights from data that might be overlooked in traditional tabular or textual representations \citep{noroozi2019multimodal, sahin2021visualizations}. By transforming raw data into interactive graphs, heat maps, dashboards, and charts, VLA could facilitate a deeper understanding of student behaviours, engagement levels, and learning progress \citep{sahin2021visualizations}. These visualisations are not only essential for educators to track and assess student performance but also empower students to engage more actively with their learning processes by enabling self-reflection \citep{vieira2018visual, verbert2020learning}. However, despite its potential, VLA faces challenges primarily due to varying levels of data visualisation literacy among users. Many students and educators lack the skills to interpret increasingly complex visual data, which can be a barrier to effectively leveraging the insights provided by VLA \citep{donohoe2020data}. 

\subsection{Scaffolding with Data Stories}

Scaffolding, a well-established educational approach, could play a crucial role in facilitating students' comprehension of VLA by breaking down complex information into manageable components. This method involves posing guiding questions and providing targeted feedback to aid students in grasping specific concepts \citep{gibbons2002scaffolding}. By promoting self-regulated learning, scaffolding encourages metacognitive processes and monitoring behaviours \citep{wen2024learning, lim2023effects}, which could potentially lead to improved comprehension when interpreting VLA. Empirical evidence supports its effectiveness in facilitating a deeper understanding and mastery of subject matter, offering contextual guidance throughout the learning process \citep{gibbons2002scaffolding, kim2018effectiveness}. 

Data storytelling has emerged as an effective scaffolding technique to enhance individuals' comprehension of complex VLA \citep{echeverria2018exploratory, pozdniakov2023teachers}. Research indicates that this approach can efficiently direct attention to key pedagogical insights \cite{echeverria2018driving, martinez2020data}, foster deep reflections, and help teachers and students understand their learning processes \cite{echeverria2018driving, fernandez2022beyond}. It is particularly beneficial for individuals with low data visualisation literacy, aiding them in efficiently interpreting information \cite{pozdniakov2023teachers}. The advantages of data storytelling are further supported by research showing its effectiveness in improving the efficiency of insight comprehension. \citet{shao2024data} found that data stories enhance information retrieval and the understanding of visual data, regardless of an individual's prior data visualisation literacy. However, it remains uncertain whether these improvements persist after the removal of storytelling elements, or if they lead to lasting comprehension of key concepts. Challenges continue in adapting and scaling data storytelling in educational settings \citep{fernandez2024data, li2024we}. Notably, research by \citet{li2024we} highlights potential integrations of data storytelling and scaffolding with emerging AI technologies, such as GenAI agents, to provide personalised support for students.

\subsection{Generative AI Agents}

The latest developments in GenAI have opened new avenues for facilitating students' comprehension of VLA \citep{chen2024beyond, li2024we}. Large language models (LLMs) such as GPT and Llama are adept at generating text-based content from prompts, which is invaluable for summarising complex datasets and visualisations, providing explanatory narratives, and crafting data stories with minimal human oversight \citep{chung2022talebrush, yan2024vizchat}. The emergence of multimodal models like GPT-4o further enhances GenAI's capability to interpret and synthesise both textual and visual data, promising more comprehensive explanations \citep{chen2024beyond, ye2024generative}. This advancement points towards potential methods that render VLA comprehension more interactive \citep{ye2024generative, li2024we}. However, simply generating descriptions of visual data might not facilitate understanding and learning \citep{Zdanovic2022}, necessitating contextually grounded explanations for effective comprehension and deeper learning insights \citep{milesi2024data, Borges2022}. A significant barrier has been the tendency of GenAI to produce well-articulated but inaccurate content, known as hallucination \citep{ji2023survey, leiser2024hill}. This can be mitigated by retrieval-augmented generation (RAG), which confines content generation to relevant material \citep{shuster2021retrieval}. By converting relevant information into vector embeddings that capture semantic meanings and retrieving this information through semantic search \citep{li2024matching}, this approach reduces hallucinations and increases domain-specific accuracy \citep{siriwardhana2023improving}. Therefore, integrating RAG with GenAI could offer real-time, contextually relevant explanations to help students understand VLA and address potential confusion \citep{yan2024vizchat}.

The development of GenAI agents, which are autonomous, adaptive entities that operate independently to achieve predefined goals without constant user input \citep{wu2023autogen}, can also contribute to enhancing students' comprehension of VLA by proactively guiding them with scaffolding questions \citep{park2023generative, oertel2020engagement}. Unlike passive agents that merely respond to queries, proactive GenAI agents can prompt students to focus on specific parts of visualisations with step-by-step guidance, similar to how an educator might ask, "What trends do you notice in this graph?" or "Why do you think this data point is an outlier?" \citep{gibbons2002scaffolding}. This structured approach could potentially support deeper understanding by breaking down complex information and offering guidance as students build their knowledge and skills \citep{kim2018effectiveness}. The educational potential of these agents aligns closely with Vygotsky's Zone of Proximal Development (ZPD), suggesting learning is most effective when students receive guidance for tasks beyond their current independent capabilities \citep{vygotsky1978mind}. As a 'more knowledgeable other,' these agents could encourage exploration and problem-solving with VLA, offering hints and scaffolded questions that help students progressively build their understanding \citep{yan2024promises, wen2024learning}. Although this method seems promising and innovative system frameworks are being developed \citep{yan2024vizchat, ma2023demonstration}, empirical evidence remains limited on whether such agents improve the effectiveness and efficiency of students' comprehension of VLA compared to conventional scaffolding approaches like data storytelling. It also remains unclear whether genuine learning occurs rather than just temporary performance improvements.
\section{Method}
\subsection{Study Contexts and Visual Learning Analytics}

\begin{figure*}[htbp]
    \centering
    \includegraphics[width=1\linewidth]{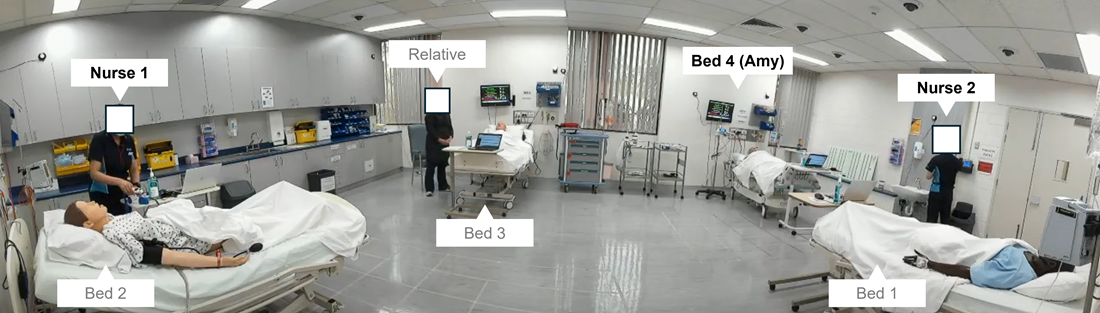}
    \caption{Picture of the healthcare simulation activity with key components labelled and identities masked.}
    \label{fig:simulation}
\end{figure*}

The current study leverages a multimodal dataset to investigate the comprehension of VLA in a healthcare simulation learning setting. The simulated learning environment (see Figure \ref{fig:simulation}), featuring four patient beds equipped with medical devices like oxygen masks and vital signs monitors, used advanced patient manikins that could simulate different heart rates and pulses, controlled by teaching staff via a tablet. The main goal of the VLA was to provide insights into students' teamwork, communication, and prioritisation skills in handling clinical emergencies during the simulation. 

The dataset used to generate the VLA (further elaborated below) included three primary types of physical and physiological data: positioning, audio, and heart rate. Students' x-y positioning data were collected using the Pozyx creator toolkit\footnote{\url{https://www.pozyx.io/creator}}, audio data were captured using wireless headsets with unidirectional microphones for multi-channel audio recording, and heart rate data were recorded using Empatica E4 wristbands.

\begin{figure*}[htbp]
    \centering
    \includegraphics[width=0.99\linewidth]{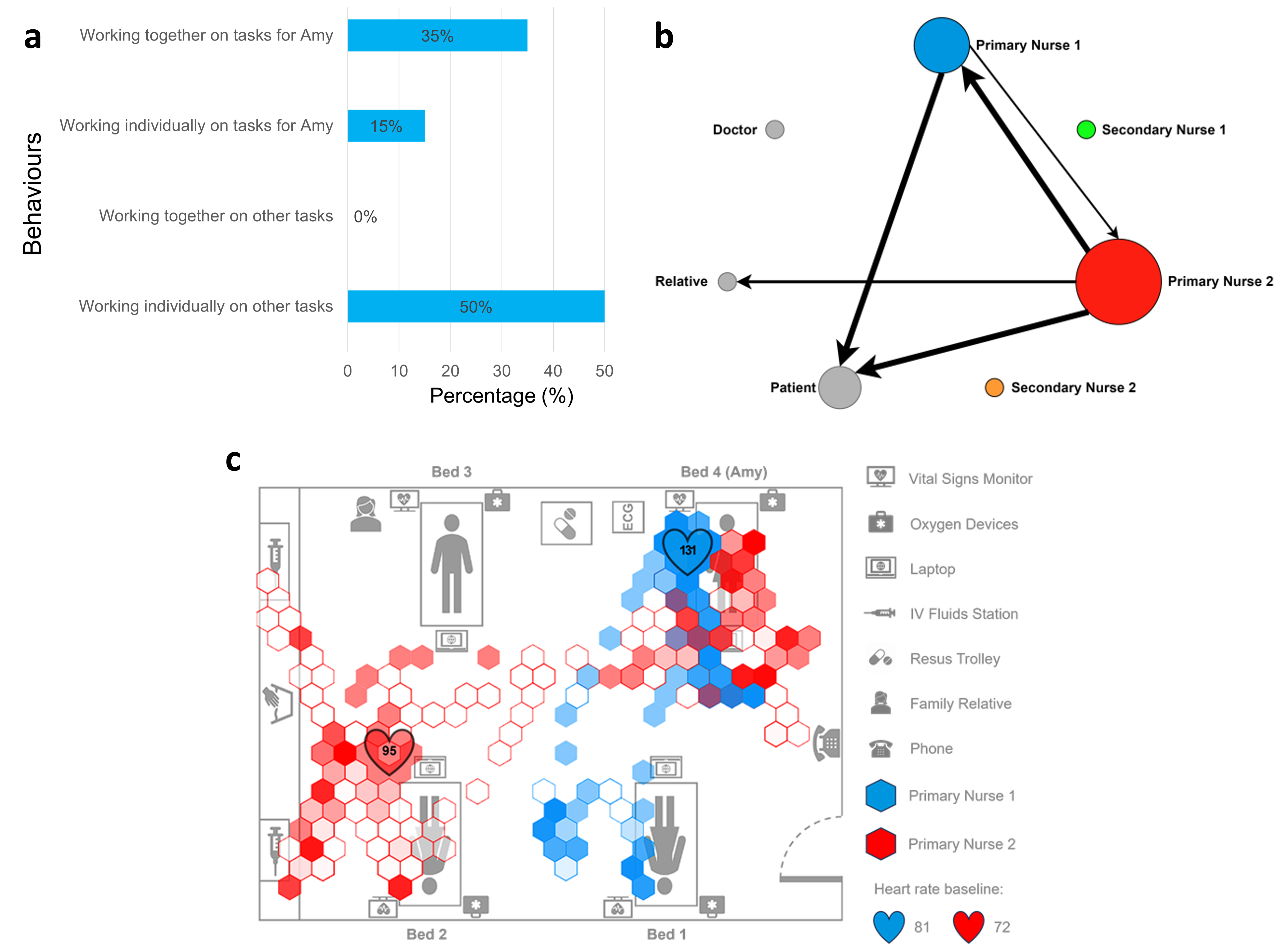}
    \caption{Visual learning analytics of teamwork in healthcare simulations: a) a bar chart showing students' four prioritisation strategies, b) a social network illustrating communication behaviours among students and other actors, and c) a ward map displaying students' physical positions (hexagon position), verbal communication durations (colour saturation), and peak heart rate locations.}
    \label{fig:visualisations}
\end{figure*}

This dataset was selected because the VLA have already been validated and utilised by healthcare educators and students for reflective purposes \citep{yan2024evidence}. As depicted in Figure \ref{fig:visualisations}, three different visualisations with increasing complexity were used. The first visualisation is a bar chart illustrating four prioritisation strategies demonstrated by students during the simulation. This type of visualisation is typically effective for comparing different categories \citep{saket2018task}. For example, the bar chart allows for a straightforward interpretation of the proportion of time each team spent on various prioritisation behaviours.

The second visualisation is a social network, or sociogram, mapping the communication behaviours among students and other actors in the simulation. Social network analysis is commonly used in collaborative learning and computer-supported cooperative work studies \citep{de2007investigating, dado2017review}. This visualisation can help students to understand interaction dynamics within the team, identifying who communicated most frequently and the direction of these communications. It can also highlight interactions with the patient, the doctor, and the relative, providing insights into the roles and engagement levels of each participant.

The third visualisation is a ward map that displays students' physical positions, verbal communication duration, and peak heart rate locations. Inspired by advanced techniques in sports analytics \citep{goldsberry2012courtvision}, this visualisation combines heatmaps \citep{gu2022complex} with specific healthcare-relevant information. The heatmap component illustrates the frequency and distribution of verbal communications, with saturated colours indicating areas of high verbal activity. Additionally, the map shows the spatial distribution of students around the simulation space, providing a comprehensive view of their physical and behavioural engagement. Furthermore, the map includes the location and value of each student's peak heart rate, indicating areas of highest physiological arousal. These visualisations collectively offer insights into prioritisation, teamwork, and communication strategies, which are challenging to comprehend without advanced visualisation literacy \citep{yan2024evidence}.

\subsection{Intervention Design: Generative AI Agents}

\begin{figure*}
    \centering
    \includegraphics[width=1\linewidth]{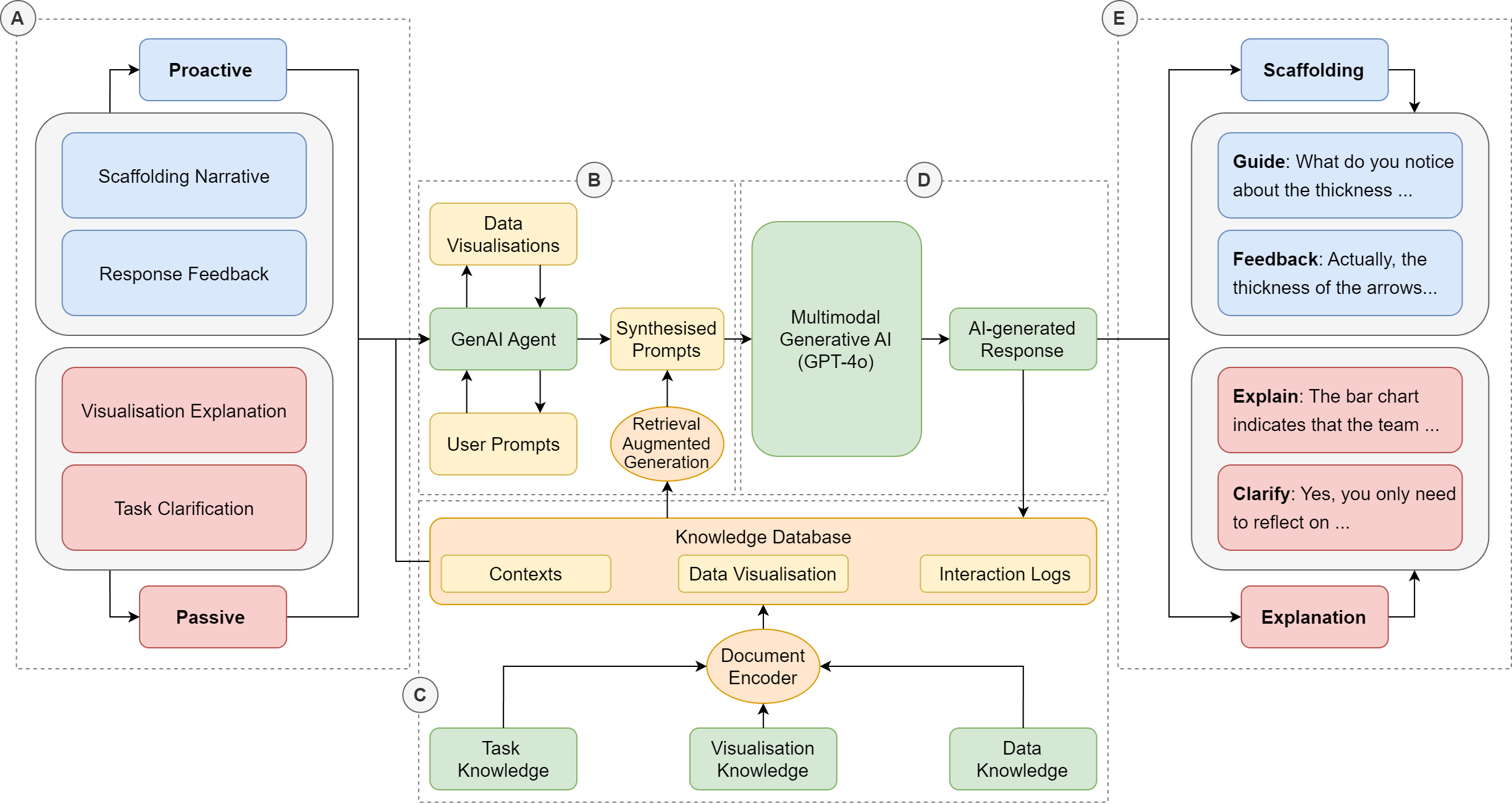}
    \caption{System design of the passive and proactive generative AI (GenAI) agents, illustrating five main components: A) unique characteristics that differentiate the passive agent from the proactive agent, B) interaction among user prompts, data visualisation, and the GenAI agent for prompt synthesis using retrieval-augmented generation, C) a knowledge database containing essential contextual information for the current task, D) generation of contextually relevant responses using multimodal GenAI, and E) examples of the resulting behaviours of the passive GenAI agent, which focuses on explanation and clarity, and the proactive GenAI agent, which provides scaffolding with guided questions and feedback.}
    \label{fig:genai}
\end{figure*}

To investigate the impact of passive and proactive GenAI agents on students' ability to comprehend the VLA, we utilised an open-sourced prototype of [Anonymised]\footnote{Hidden for double-blinded review}. This prototype integrates multimodal GenAI (e.g., OpenAI GPT-4o\footnote{\url{https://openai.com/index/hello-gpt-4o/}}), RAG frameworks (e.g., LangChain), and agentic system designs (e.g., Autogen\citep{wu2023autogen}) to ensure the accuracy and contextual relevance of information provided by the GenAI agents. Multimodal GenAI processes both graphical and textual data, tailoring AI-generated responses specifically to the visualisations participants inquire about. RAG incorporation ensures GenAI agents access relevant and accurate knowledge for context-specific prompts \citep{lewis2020retrieval,gao2023retrieval}. Agentic systems \citep{park2023generative} enable GenAI agents to address irrelevant responses generated by incomplete user prompts \citep{white2023prompt}. As shown in Figure \ref{fig:genai}, the design of the passive and proactive GenAI agents includes five main components: i) agent characteristics, ii) prompt integration, iii) knowledge database, iv) response generation, and v) exemplar behaviours.

\subsubsection{Agent characteristics.} The core interaction styles distinguish the passive and proactive GenAI agents. Passive GenAI agents are reactive, waiting for student-initiated queries before responding \citep{ma2023demonstration, yan2024vizchat}. They are characterised by their ability to deliver precise and contextually relevant information without initiating unsolicited interactions. In contrast, proactive GenAI agents are more engaging and interactive \citep{park2023generative, oertel2020engagement}. They not only respond to student queries but also guide students through the visualisations with structured narratives and scaffolding questions (crafted by educators and researchers; Appendix \ref{appendix:scaffold}). This dual approach allows for a comparative analysis of how different interaction styles impact user understanding and insight extraction from VLA.

\subsubsection{Prompt integration.} Prompt integration is critical for both passive and proactive GenAI agents. For passive agents, the process starts with student-initiated queries, which are processed to generate contextually relevant responses. This involves interpreting the natural language input and matching it with the visualisation and the knowledge database using RAG. Proactive agents, however, go a step further by synthesising prompts based on user interactions and a pre-defined scaffolding narrative consisting of guiding questions that facilitate the exploration of the visualisations. They generate prompts to guide students through the visualisations, helping them understand the data more structurally and providing feedback based on students' responses. This is achieved through a combination of pre-stored knowledge and real-time data interpretation, ensuring accurate and tailored responses.

\subsubsection{Knowledge database.} The knowledge database is the backbone of the GenAI agents, providing the necessary contextual information for accurate and relevant responses. Built using LangChain\footnote{\url{https://www.langchain.com/}} and Chroma\footnote{\url{https://www.trychroma.com/}}, the database is populated with task-specific materials, descriptions, and detailed explanations of visualisation components. These materials are directly drawn from the learning design documents of the healthcare simulation unit and peer-reviewed publications describing each visualisation [Anonymised]. Textual materials are converted into vector embeddings using OpenAI's embedding model (text-embedding-ada-002), allowing for efficient semantic searches. This enables the agents to retrieve pertinent information based on the cosine similarity of the embeddings \citep{gao2023retrieval}. The knowledge database is dynamic, evolving with each user interaction. New information from student-agent conversations is continuously ingested, enhancing the personalisation and relevance of subsequent interactions.

\subsubsection{Response generation.} Response generation in both passive and proactive GenAI agents utilises the multimodal capabilities of GPT-4o. For passive agents, the process involves generating responses based on user queries by accessing the knowledge database and retrieving relevant information using RAG. The responses are designed to be precise and contextually relevant, ensuring that students receive accurate answers to their questions. Proactive agents follow pre-defined scaffolding narratives with guiding questions. Each response contains feedback on students' prior responses and asks students to either elaborate on the current response or move on to the next guiding question. This involves performing semantic searches across the vector embeddings to identify relevant text segments, assessing the correctness of students' responses, and generating encouraging or corrective feedback accordingly.

\subsubsection{Exemplar behaviours.} 
\begin{figure*}
    \centering
    \includegraphics[width=1\linewidth]{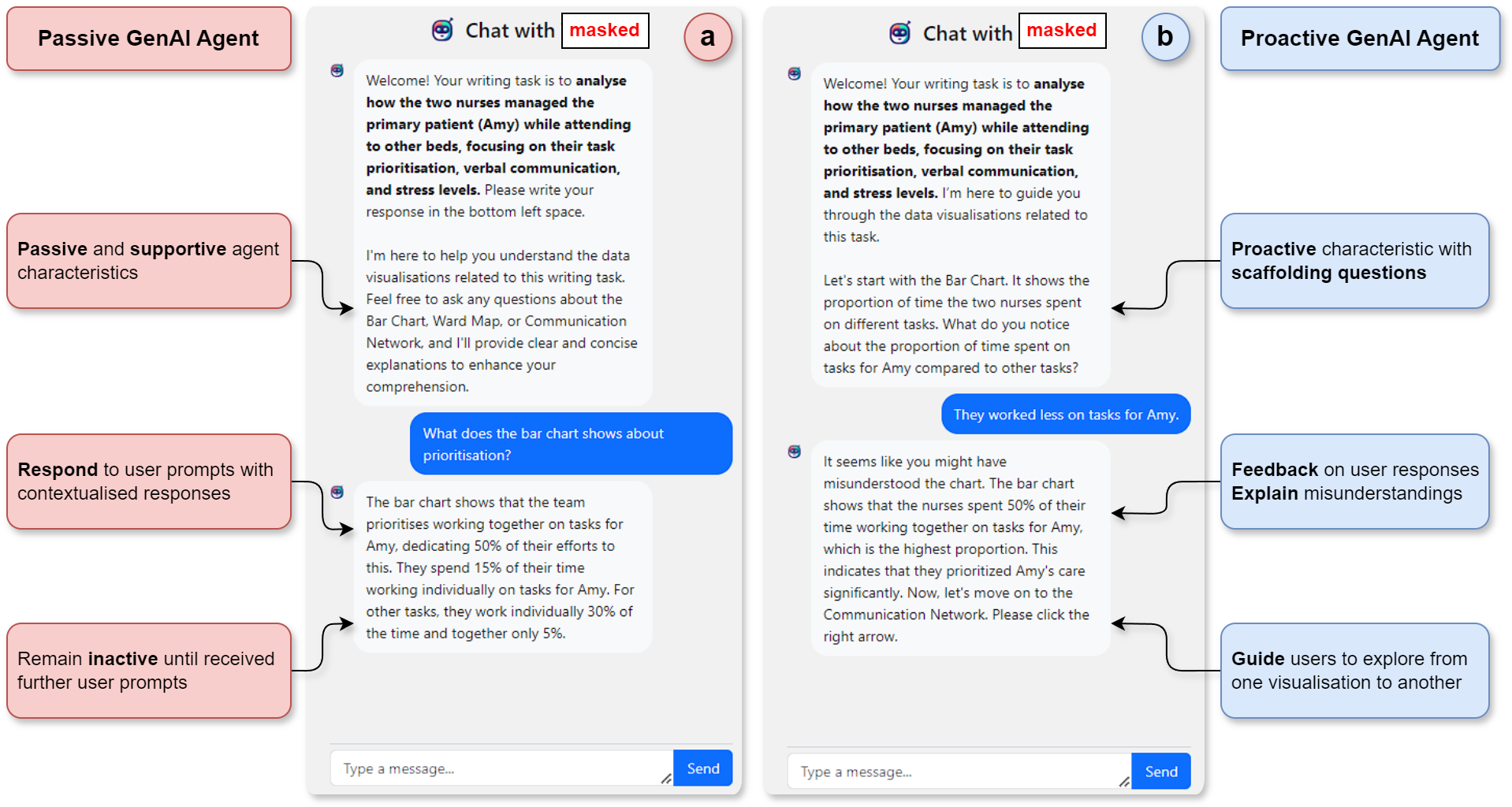}
    \caption{Example interactions with the two generative AI (GenAI) agents, illustrating the key characteristics of a) the passive GenAI agent and b) the proactive GenAI agent.}
    \label{fig:genai-examples}
\end{figure*}

The behaviours of the passive and proactive GenAI agents are exemplified by their interaction styles (Figure \ref{fig:genai-examples}). Passive agents focus on explanation and clarity, providing students with straightforward answers to their queries. For example, when asked about the representation of a bar chart, a passive agent retrieves and presents the relevant information from the knowledge database. Proactive agents, however, offer scaffolding with guided questions and feedback. They actively engage students by asking clarifying questions, offering step-by-step explanations, and facilitating the exploration of the visualisations within a scaffolding framework (e.g., a set of guiding questions targeting each element of the visualisations) \citep{xun2004conceptual}. For example, to facilitate students' understanding of a social network, a proactive agent might start by asking questions about the thickness of the edges connecting two nodes. After confirming students have a clear understanding through their responses, it will then move on to questions about the direction of the edges. This scaffolding approach aims to structurally enhance comprehension and cultivate the ability to comprehend data visualisations.

\subsection{Intervention Design: Data Stories}

\begin{figure*}
    \centering
    \includegraphics[width=1\linewidth]{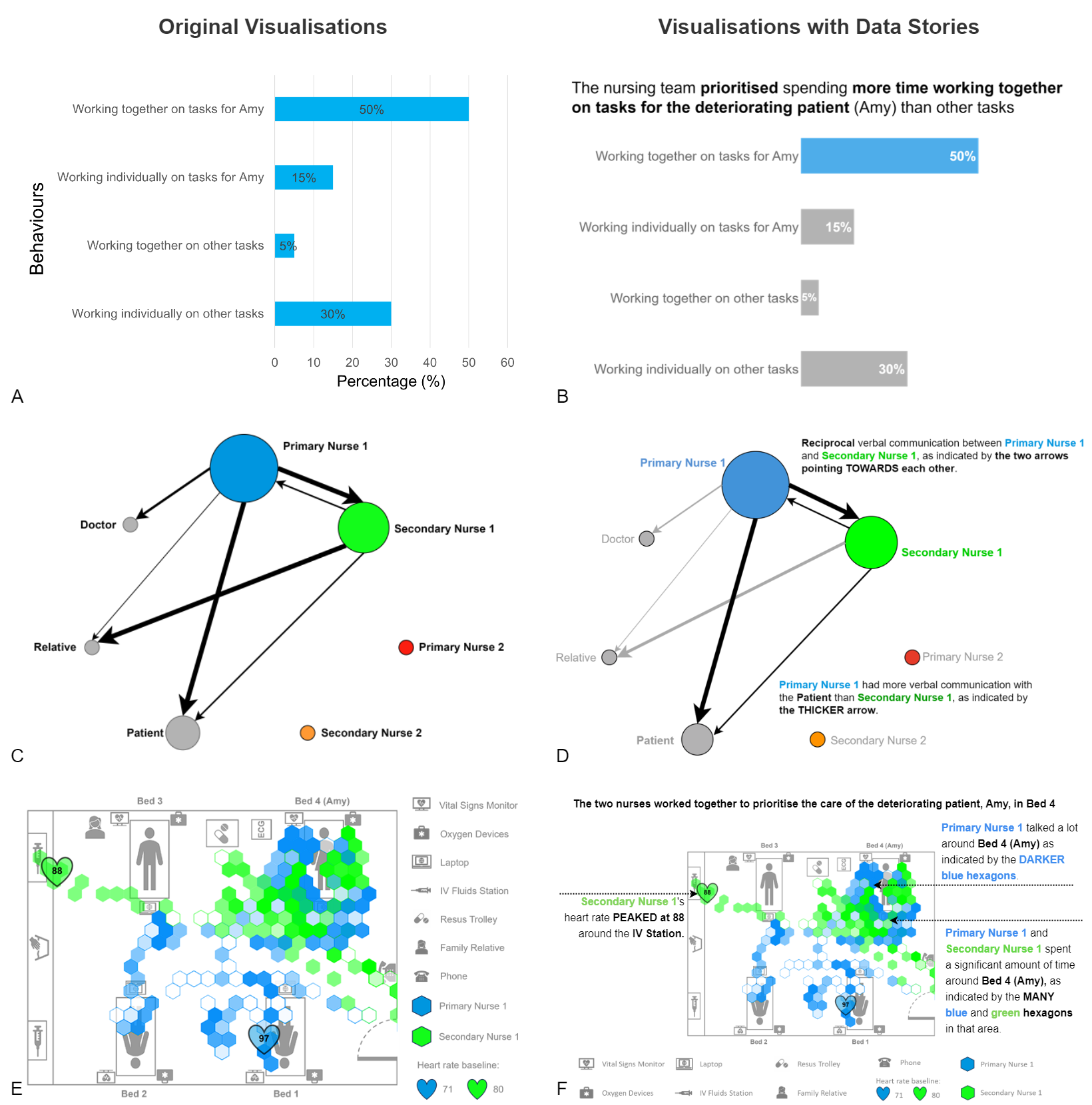}
    \caption{Visual learning analytics infused with data storytelling elements. }
    \label{fig:stories}
\end{figure*}

To reliably create data visualisations enhanced with storytelling elements, we identified key techniques based on the framework by \citet{Zdanovic2022} and followed a structured three-phase process: exploring the data, creating the visualisation, and telling the story. Key elements included choosing the right visualisation technique to match storytelling goals \citep{Knaflic2015}, simplifying visuals by eliminating clutter to enhance readability \citep{tufte2001visual, Heer2012interactive}, directing attention using bold lines or contrasting colours for key insights \citep{Knaflic2015, Ware2019Information}, adding annotations for context \citep{Hullman2013}, and crafting explanatory titles that guide viewers \citep{pozdniakov2023teachers, Kong2018Frames}. As shown in Figure \ref{fig:stories}, our process did not involve changing chart types, maintaining complexity progression from bar charts to social network visualisations, and custom-made ward maps. Initially, one researcher drafted the data stories, which were then reviewed and refined collaboratively by researchers with expertise in data storytelling, ensuring alignment with research objectives. This iterative approach ensured thoughtfully developed visual elements, adhering to Zdanovic's framework for effective storytelling.

\subsection{Experiment Design and Procedure}

This randomised controlled study was designed following gold standards for effectiveness research \citep{hariton2018randomised}. We conducted a mixed-design (3x3) experiment, using both within-subject and between-subject comparisons to evaluate the effectiveness of data storytelling, passive GenAI agents, and proactive GenAI agents in enhancing insight extraction from VLA. The study was conducted using Qualtrics\footnote{\url{https://www.qualtrics.com/}}, an online platform for designing and conducting experiments, integrated with a custom-developed website. Participants were recruited from Prolific\footnote{\url{https://www.prolific.co/}}, compensated £8, and the study was structured to be completed within one hour. Ethics approval was obtained from [Anonymised] University (Project Number: Anonymised). Informed consent was secured from each participant.

Participants were randomly assigned to one of three intervention conditions: passive GenAI agents, proactive GenAI agents, or data storytelling. Passive GenAI agents responded to user queries without initiating further dialogue or guidance, while proactive GenAI agents guided users with scaffolding questions. The data storytelling condition presented visualisations within a narrative format, highlighting key insights and trends. Each participant interacted with their assigned condition during the intervention phase to support insight extraction from visualisations. The effectiveness of each method was evaluated through pre-, during, and post-intervention assessments.

\begin{figure*}[htbp]
    \centering
    \includegraphics[width=1\linewidth]{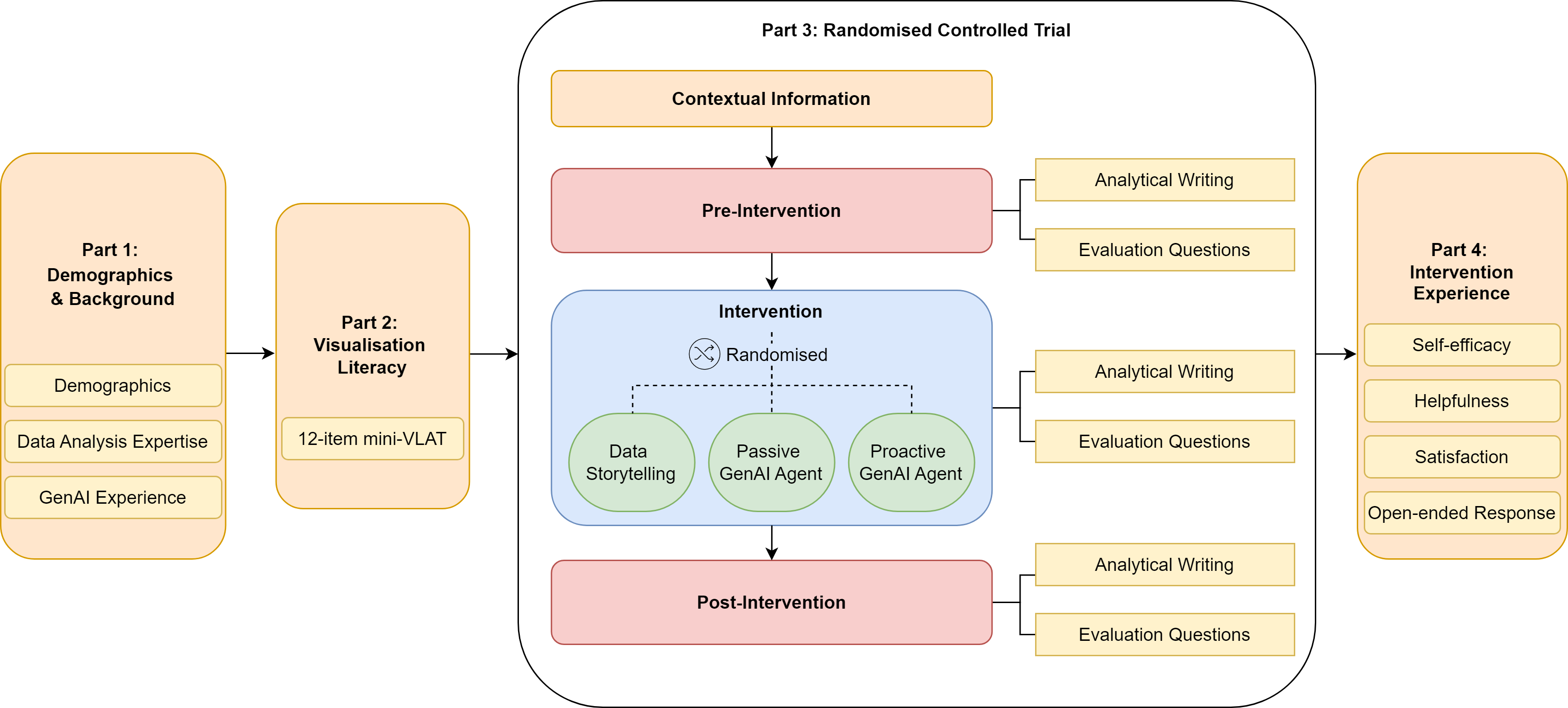}
    \caption{Experiment design and procedure: Detailed breakdown of participant flow through demographics and background questionnaires, visualisation literacy assessment, the randomised controlled trial, and intervention experience questionnaire.}
    \label{fig:procedure}
\end{figure*}

The study followed a structured procedure to ensure consistency across conditions, comprising five main parts: i) the participants received an introduction to the study and completed demographic and background questions on Qualtrics; ii) they answered 12 items from the mini-Visualisation Literacy Assessment Test (mini-VLAT)\citep{pandey2023mini} to measure their visualisation literacy; iii) the participants were directed to our custom platform to complete three analytical writing tasks, each followed by six evaluation questions (in Qualtrics) to assess their understanding of the VLA; and iv) finally, the participants completed a questionnaire to provide feedback on their experience with the assigned intervention method. Figure \ref{fig:procedure} illustrates the experiment design and procedure. Details are elaborated below.

\subsubsection{Part 1: Demographics and background questions}

Participants were initially queried on demographic information, including self-reported gender and region. Subsequently, we assessed participants' self-reported experience in data analysis and GenAI tools. For data analysis, participants rated their experience on a scale from \textit{"None - I have little to no experience with data analysis"} to \textit{"Expert - I have extensive experience and can perform data analysis tasks confidently."} Similarly, for GenAI tools, participants rated their experience from \textit{"None - I have no prior experience with generative AI tools"} to \textit{"Expert - I have deep expertise and often develop or customise generative AI tools for specialised purposes."}

\subsubsection{Part 2: Visualisation literacy}

The 12-item mini-VLAT was used to assess participants' visualisation literacy \citep{pandey2023mini}. This instrument was selected due to its demonstrated reliability and validity, comparable to the full 53-item VLAT \citep{lee2016vlat} while offering a more concise measure of visualisation literacy. These features made it particularly suitable for our extended study, as it minimised cognitive load on participants. The scores were corrected for guessing using the correction-for-guessing formula \citep{thorndike1991measurement}:

\begin{equation}
CS = \textit{R} - \textit{W} / (\textit{C} - 1)
\end{equation}

In this formula, \textit{CS} represents the final score corrected for guessing behaviours, \textit{R} represents the number of correctly answered items of a given participant, \textit{W} represents the number of wrongly answered items, and \textit{C} represents the number of choices per item, which is four for the mini-VLAT.

\subsubsection{Part 3: Randomised controlled trial}

The randomised controlled trial consisted of four main activities: i) reading contextual information about the visualisations, ii) all participants completing the same baseline analytical writing activity (\textit{Pre-intervention}) where no intervention was provided, followed by six evaluation questions, iii) each participant being randomly assigned to one of the three intervention conditions (\textit{Intervention}), having access to either data storytelling, passive GenAI agents, or proactive GenAI agents, followed by six evaluation questions, and iv) all participants completing another, yet different, baseline activity (\textit{Post-intervention}) with the intervention removed, followed by six evaluation questions. 

The visualisation format remained consistent across the pre-intervention phase, the intervention phase, and the post-intervention phase (e.g., bar chart, social network, and ward map), although the data and insights contained within these visualisations varied across these different phases. Likewise, the six evaluation questions maintained the same format, with the questions and answers tailored to the insights for each activity. Details of the contextual information, analytical writing tasks, and evaluation questions are elaborated below:

\textbf{Contextual information.} Participants were provided with a picture (Figure \ref{fig:simulation}) and a description of a healthcare simulation where two nursing students took over care of four manikin patients (Beds 1-4) during their shift. Each patient required at least two tasks, with the primary focus on the deteriorating patient in Bed 4 (Amy), while the needs of other patients (Beds 1-3) also required attention. An actor played a relative of the patient in Bed 3, frequently distracting the nurses. The simulation aimed to practice teamwork, communication, and prioritisation skills. By the end, students were expected to i) demonstrate a structured approach to patient assessment and management, ii) recognise and respond to early signs of deterioration, and iii) contribute to effective teamwork. This background was essential for participants to engage in the subsequent writing tasks effectively.

\begin{figure*}[htbp]
    \centering
    \includegraphics[width=1\linewidth]{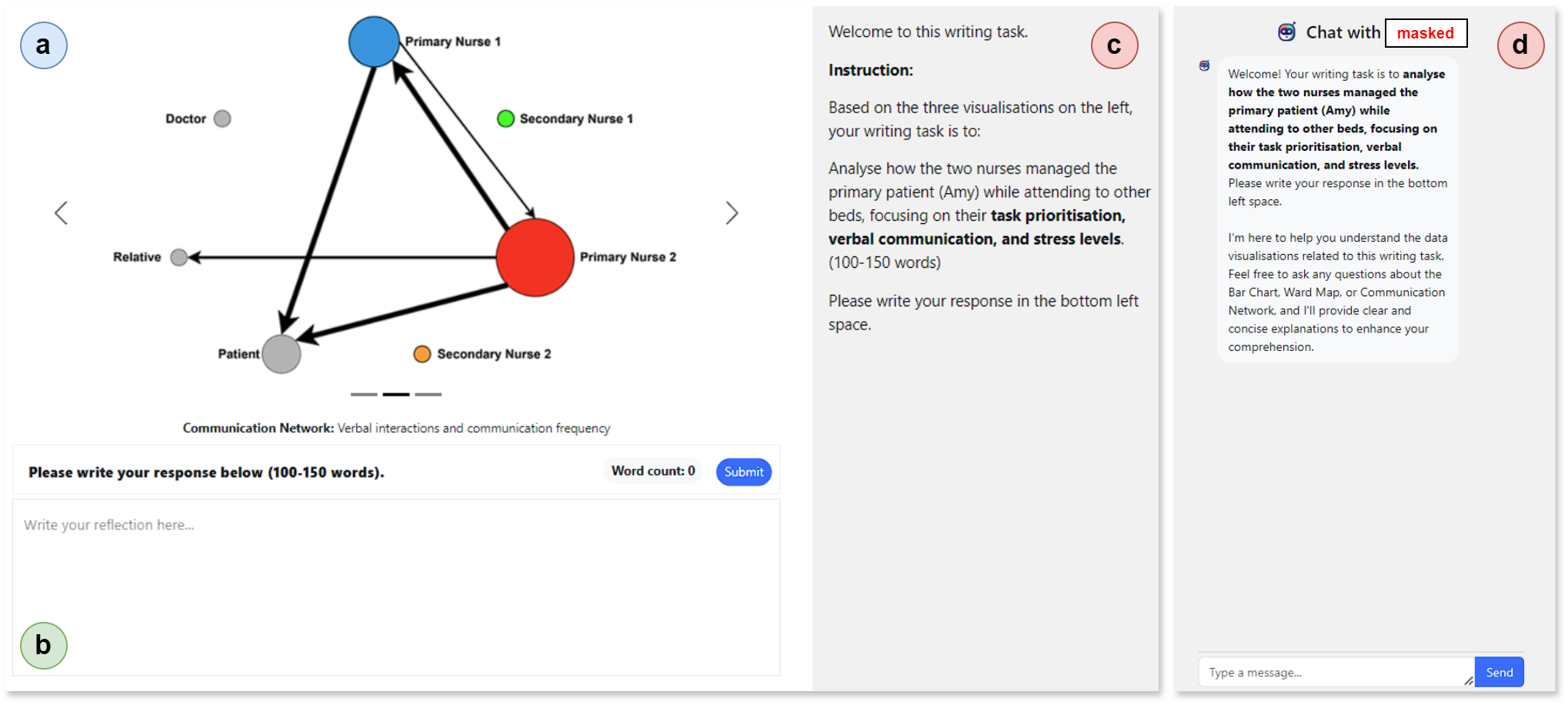}
    \caption{Custom-developed website for the analytical writing task. a) a display component for the visualisations, b) a writing space, c) task instructions used in the pre-intervention, post-intervention baselines, and data storytelling condition, and d) a chat function with GenAI agents for the passive or proactive conditions.}
    \label{fig:website}
\end{figure*}

\textbf{Analytical writing tasks.} Participants were directed to a custom-developed website (Figure \ref{fig:website}) to complete an analytical writing task. They were asked to analyse three visualisations (bar chart, social network, and ward map) and write a 100-150 word response on \textit{how the two nurses managed the primary patient (Amy) while attending to other beds, focusing on their task prioritisation, verbal communication, and stress levels}. The website was designed with simplicity in mind \citep{stone2005user}, ensuring easy navigation and task completion. 

The website had three key components. First, the display component (Figure \ref{fig:website}a) presented the visualisations one at a time to avoid information overload and visual clutter \citep{ellis2007taxonomy}. Second, the writing space (Figure \ref{fig:website}b) allowed participants to articulate their analysis, synthesising information from the visualisations. Lastly, the instructions component (Figure \ref{fig:website}c) provided standard task instructions for the pre-intervention and post-intervention baselines and the data storytelling condition. In the passive and proactive GenAI conditions, participants interacted with GenAI agents via a chat function for additional support and guidance (Figure \ref{fig:website}d). 

This design helped participants focus on the analytical writing task without distractions, facilitating clear and effective exposure to each intervention condition and providing opportunities to develop their ability to comprehend data visualisations.

\textbf{Evaluation questions.} After each analytical writing task, participants were asked to complete six multiple-choice questions (four choices per question; two for each visualisation; Appendix \ref{appendix:evaluation}). They complete these questions without any form of support in Qualtric instead of on the custom-built website. These questions were designed to evaluate their ability to comprehend data visualisations. Structured based on the first two levels of Bloom's taxonomy, knowledge (Level 1) and comprehension (Level 2) \citep{bloom1984bloom}, these questions directly align with common practices in visualisation research \citep{arneson2018visual,mnguni2016assessment}. For the knowledge questions, participants were required to identify specific data points or patterns in the visualisations, such as pinpointing the prioritisation behaviour that two nurses spent the most time on from a bar chart. These questions assessed their ability to retrieve pertinent information. For the comprehension questions, participants had to interpret and derive meaningful insights, such as comparing the spatial and verbal activities between two nurses through the ward map. These questions evaluated their ability to interpret multiple insights and identify inconsistencies or misleading information. Higher levels of Bloom's taxonomy, such as application (Level 3), were not addressed, as the study involved online participants with limited contextual knowledge.

Two researchers familiar with the learning contexts and visualisations designed the six evaluation questions. A third researcher validated this design to ensure its applicability across all three sets of visualisations: pre-intervention, intervention, and post-intervention. Any discrepancies were resolved through discussion, leading to a consensus on the final evaluation questions. These questions ensured alignment with the data insights in the visualisations and the appropriate Bloom's taxonomy levels. While the order of the questions remained consistent across all phases, the order of the answer choices was randomised to minimise bias and improve reliability and validity \citep{aera2014}. A fifth option, 'I am not sure,' was added to each question to reduce guessing, following recommendations in mini-VLAT and VLAT \citep{lee2016vlat, pandey2023mini}. The time taken to answer each question correctly was measured to evaluate the efficiency of insights extraction, while the accuracy rate was recorded to provide insights into the effectiveness of comprehension (further elaborated in Section \ref{sec:method-analysis}).

\subsubsection{Part 4: Intervention experience}

For the last part of the study, participants were asked to evaluate their experiences with the intervention (e.g., data storytelling, passive GenAI agents, or proactive GenAI agents) through three single-choice items and one open-ended question. Each item was rated on a five-point Likert scale ranging from strongly disagree (1) to strongly agree (5). Table \ref{tab:experience} shows the items. Although the internal reliability of these single-item items can not be estimated, they have shown evident reliability in capturing individuals' attitudes and beliefs \citep{wanous1996estimating}. The open-ended question targeted further qualitative elaborations -- \textit{"Please describe your experience with the intervention. To what extent did it enhance or diminish your ability to interpret the data visualisations?"}

\begin{table*}[htbp]
\centering
\caption{Intervention Experience Items}
\label{tab:experience}
{
\renewcommand{\arraystretch}{2}
\begin{tabular}{ll}
\hline
\textbf{Dimension} & \textbf{Item}\\
\hline
Self-efficacy & I feel more confident in my ability to interpret data visualisations after participating in the intervention.\\
\hline
Helpfulness & The intervention helped me better understand the data visualisations.\\
\hline
Satisfaction & I am satisfied with my overall experience during the intervention.\\
\hline
\end{tabular}
}
\end{table*}

\subsection{Participants}

We conducted a priori power analysis for a mixed-design study with three between-subject conditions and three within-subject repeated measures using G*Power \citep{faul2009statistical}. The analysis indicated that a minimum sample size of 108 participants (36 per condition) was needed to detect a medium effect size (0.25) with 80\% power at a 0.05 significance level. To account for an anticipated attrition rate of 15\% to 20\% due to potential data issues, particularly given the one-hour duration of the online study, we aimed to recruit 150 participants. Participants were required to be fluent in English and use a laptop or desktop to ensure consistent visualisation displays. We specifically recruited higher education students with healthcare, medical, or nursing backgrounds to control for familiarity with the healthcare simulation context of the VLA. 

Before the main study, a pilot study with 27 participants (9 per condition) was conducted to refine the Qualtrics survey for clarity and ease of use. We also tested the custom-developed website to ensure timely responses from both passive and proactive GenAI agents. Data from the pilot study were excluded from the final analysis due to subsequent improvements in the study flow, such as mandatory reading time for each task instruction and improved clarity of several context description sentences.

For the main study, participants had to complete all three writing tasks, essential for internalising their understanding of data visualisations. Incomplete responses, including failure to complete any writing task or the entire survey, were deemed invalid. Out of the 141 responses received, 24 were invalid due to incomplete writing tasks, resulting in a final sample of 117 valid responses. Specifically, 36 participants completed the data storytelling condition, 41 completed the passive GenAI agents condition, and 40 completed the proactive GenAI agents condition.

The participants came from six different regions, with the majority coming from North America/Central America (51), Europe (28), and Africa (25). The remainder were from Australia (5), South America (4), and other regions (4). They identified as female (68), male (48), and non-binary (1). Regarding experience with data analysis, most participants were at an intermediate level (50), with a moderate level of experience. Others were beginners (36), advanced users (17), had no experience (12), or were experts (2). In terms of familiarity with GenAI tools, most participants were at an intermediate level (63), having used the tools on several occasions. Others were beginners (34), advanced users (15), experts (4), or had no prior experience (1). 

\subsection{Analysis}
\label{sec:method-analysis}
The analysis aimed to address three research questions (RQs) regarding the impacts of different interventions on enhancing participants' effectiveness and efficiency in extracting insights from VLA. We first calculated a set of metrics related to the RQs using the responses and times for six evaluation questions for each task (Appendix \ref{appendix:evaluation}). These metrics were calculated for both within- and between-subject analyses, including the three intervention phases, specifically, the pre-intervention baseline (Pre), the intervention phase, and the post-intervention phase (Post), as well as for the three intervention conditions, namely data storytelling (DS), passive GenAI agents (GAI), or proactive GenAI agents (DSAI). In particular, the following metrics were calculated:
\begin{itemize}
\item \textbf{Correct Score}: To investigate the effectiveness of insights extraction from data visualisations (RQ1-2), we computed the correct score for the six evaluation questions for each condition by summing the total number of questions that participants answered correctly (each score ranges from 0 to 6), resulting in three correct scores for each participant (Pre\_score, Intervention\_score, and Post\_score). The correction-for-guessing formula \citep{thorndike1991measurement} was applied to account for guessing in these multiple-choice questions (with four options).
\item \textbf{Success Time}: To investigate participants' efficiency in data insights extraction (RQ1-2), we summed the total time they spent on answering the six evaluation questions for each condition in seconds and then divided by the corresponding correct score for that given condition \citep{shao2024data}, resulting in three success time metrics for each participant (Pre\_time, Intervention\_time, and Post\_time). 
\end{itemize}
In addition to the above two metrics, visualisation literacy (mini-VLAT) was calculated separately for each participant to address RQ3. This calculation was done by summing the number of correct responses. Specifically, visualisation literacy ranges from 0 to 12. For both literacy scores, the correction-for-guessing formula \citep{thorndike1991measurement} was applied to account for guessing behaviours. All calculations and analyses were conducted in Python using packages including NumPy, SciPy, and Statsmodels. The following sections elaborate on the analysis for each RQ. 

\subsubsection{Preliminary Analysis: Learning Effects and Comparability} 

Before addressing each research question, we first assessed whether learning effects exist due to the repeated design, which is critical for attributing any observed improvements to the interventions instead of repeated exposure \citep{charness2012experimental}. We conducted a mixed linear model analysis with correct scores (Pre\_score, Intervention\_score, and Post\_score) as the dependent variable. We included phase (pre-intervention, intervention, post-intervention) as the fixed effect and incorporated each participant as a random effect to account for variability within subjects. We analysed score changes from pre-intervention to post-intervention phases across all conditions to identify any improvements that could be attributed to repeated task exposure rather than intervention effects.

We also conducted a preliminary analysis to ensure that the three intervention conditions were comparable in terms of participants' visualisation literacy (mini-VLAT), data analysis, and GenAI expertise (demographics ranging from 1--None to 5--Expert). This preliminary analysis is essential to ensure the findings for the subsequent analyses are robust and reliable \citep{charness2012experimental}. We used the Kruskal-Wallis test, a non-parametric test that compares the medians of more than two independent groups, to assess whether there were statistically significant differences in participants' visualisation literacy, data analysis, and GenAI expertise \citep{mckight2010kruskal}. A non-significant finding for each of these metrics is critical to confirm the comparability of the three intervention conditions. Additionally, boxplots were drawn to visualise the sample distribution for each condition.

\subsubsection{RQ1: Within-subject Analysis} 

To address RQ1, we performed within-subject analyses to examine changes in participants' effectiveness and efficiency in extracting insights from data visualisations across three repeated measures for different intervention conditions. First, we utilised the Friedman test \citep{sheldon1996use}, a non-parametric test suitable for detecting differences in scores across multiple conditions. This test was chosen because of the repeated measures design and the ordinal nature of the data, as well as violations of normality in data distribution identified using the Shapiro-Wilk W-test. The Friedman test compared median correct scores (Pre\_score, Intervention\_score, and Post\_score) and median success times (Pre\_time, Intervention\_time, and Post\_time) within participants across the three study phases. The effect size was calculated using Kendall's W \citep{pereira2015overview}. Following significant results from the Friedman test, we conducted pairwise comparisons using Wilcoxon signed-rank tests \citep{wilcoxon1970critical}. This non-parametric test compared two related samples to determine whether their population mean ranks differ, making it robust for handling non-normally distributed data. This allowed us to identify which specific conditions differed significantly. The Holm-Bonferroni method was used to adjust for multiple comparisons, with an initial alpha of 0.05. Effect sizes for these tests were calculated using Rank-Biserial correlation to assess the magnitude of observed differences \citep{kerby2014simple}.

\subsubsection{RQ2: Between-subject Analysis}

To address RQ2, we conducted between-subject analyses to compare the effectiveness and efficiency of different intervention conditions across participants. We began with the Kruskal-Wallis test, a non-parametric method suitable for this analysis due to observed violations of normality, especially for Post\_score. This test compared the correct scores (Pre\_score, Intervention\_score, and Post\_score) and success times (Pre\_time, Intervention\_time, and Post\_time) across different intervention conditions to assess statistically significant differences between the conditions. The effect size was calculated using eta-squared. When the Kruskal-Wallis test indicated significant differences, we used the Mann-Whitney U test to explore differences between pairs of independent groups further. This non-parametric test is appropriate for non-normally distributed data and assesses whether two independent samples come from the same distribution. The Holm–Bonferroni method was applied to adjust for multiple comparisons with an initial alpha of 0.05. We also calculated effect sizes for these tests to quantify the strength of observed differences using the Rank-Biserial correlation \citep{kerby2014simple}.

\subsubsection{RQ3: Regression Analysis}

For RQ3, we used ordinary least squares (OLS) regression to examine the relationship between intervention conditions and visualisation literacy on participants' effectiveness and efficiency in extracting insights from data visualisations. This analysis considered both main effects (e.g., intervention conditions) and interaction effects (e.g., intervention conditions and visualisation literacy), holding other variables constant. The pre-intervention baseline (Pre) was excluded to focus on the effects during and after the intervention. First, we standardised the mini-VLAT scores by subtracting the mean and dividing by the standard deviation, ensuring normal distribution. Outliers for scores (Intervention\_score and Post\_score) and time-related metrics (Intervention\_time and Post\_time) were removed using the interquartile range (IQR) method, and data were transformed using the Box-Cox transformation to stabilise variance and normalise data distribution. For each metric (Intervention\_score, Intervention\_time, Post\_score, Post\_time), we constructed a formula to investigate the effects of visualisation literacy (mini-VLAT) across all three intervention conditions. The formula included an intercept (\(\beta_0\)), main effects for the GAI condition (\(\beta_1\)) and the DSAI condition (\(\beta_2\)), visualisation literacy (\(\beta_3\)), and interaction terms between intervention conditions and visualisation literacy (\(\beta_4\) and \(\beta_5\)), with the DS condition as the reference. We fitted the models using the OLS function from the Statsmodels library. The assumptions were validated through several methods. Linearity was assessed by plotting predicted versus observed values. The normality of residuals was checked using the Shapiro-Wilk test and QQ plots. Homoscedasticity was evaluated using the Breusch-Pagan test, and the independence of residuals was assessed using the Durbin-Watson test. All assumptions were met.

\begin{equation}
\begin{aligned}
\text{{Score or Time}} &= \beta_0 + \beta_1 \times \text{{condition [GAI]}} + \beta_2 \times \text{{condition [DSAI]}} + \beta_3 \times \text{{mini-VLAT}}\\ 
&\quad + \beta_4 (\text{{condition [GAI]}} \times \text{{mini-VLAT}}) + \beta_5 (\text{{condition [DSAI]}} \times \text{{mini-VLAT}})
\end{aligned}
\end{equation}


\section{Results}

\subsection{Preliminary Analysis -- Learning Effects and Intervention Comparability}

The preliminary analysis of the phase effects demonstrated no significant change in scores from the intervention to the post-intervention phase (\(\beta = 0.120, z = 0.948, p = 0.343\)), indicating minimal learning effects due to repeated exposure. However, a significant decrease was observed from pre-intervention to the intervention phase (\(\beta = -1.359, z = -10.769, p < 0.001\)), which aligns with expected baseline shifts as participants engage with interventions. The group variance (\( \text{Group Var} = 0.441 \)) captured variance across individual differences not attributable to fixed effects. These findings suggest that any observed improvements in subsequent analyses can be attributed to specific intervention effects rather than learning effects and familiarisation with the task. 

\begin{figure}
    \centering
    \includegraphics[width=1\linewidth]{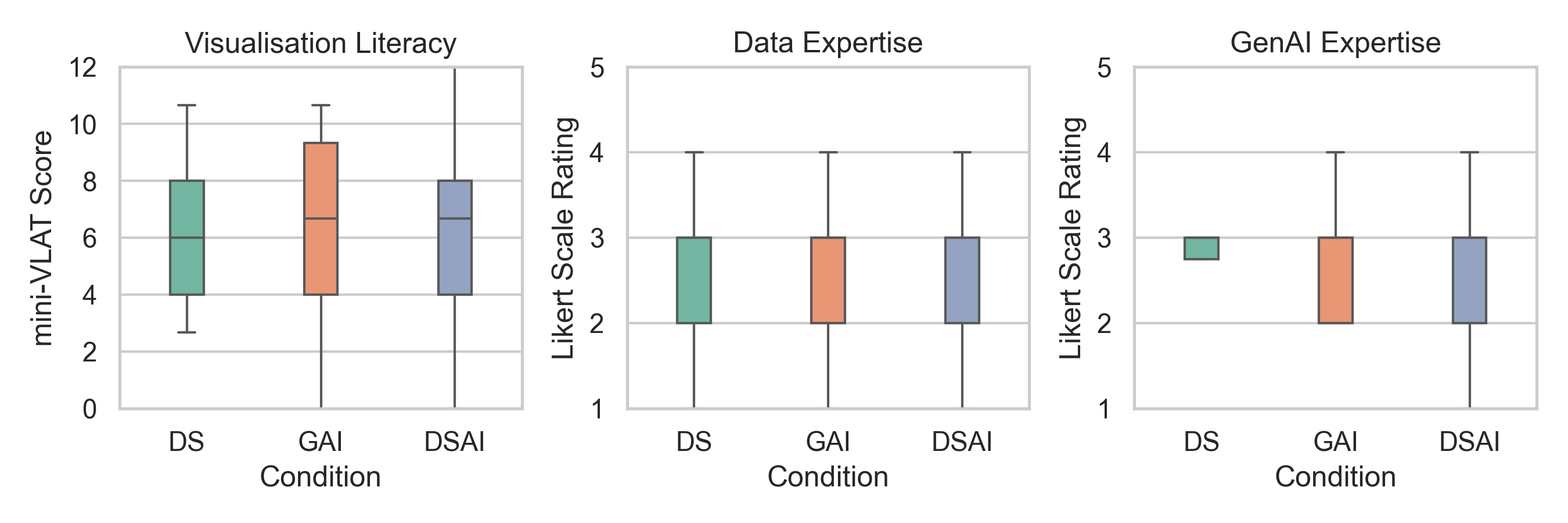}
    \caption{Comparison of visualisation literacy, data analysis, and generative AI expertise across intervention conditions.}
    \label{fig:rq0}
\end{figure}

To ensure the comparability of the three intervention conditions, we conducted a preliminary analysis evaluating participants' visualisation literacy (measured by mini-VLAT), data analysis expertise, and GenAI expertise. The Kruskal-Wallis test results indicated no significant differences in visualisation literacy between the three conditions: DS (Median = 6.0, IQR = 4.0), GAI (Median = 6.67, IQR = 5.33), and DSAI (Median = 6.67, IQR = 4.0), \(H(2) = 0.49, n = 117, p = .784\). Similarly, data expertise scores did not significantly differ between the conditions: DS (Median = 3.0, IQR = 1.0), GAI (Median = 3.0, IQR = 1.0), and DSAI (Median = 3.0, IQR = 1.0), \(H(2) = 1.02, n = 117, p = .601\). For GenAI expertise, no significant differences were found across the conditions: DS (Median = 3.0, IQR = 0.25), GAI (Median = 3.0, IQR = 1.0), and DSAI (Median = 3.0, IQR = 1.0), \(H(2) = 1.42, n = 117, p = .491\). These non-significant findings in visualisation literacy, data expertise, and GenAI expertise indicate that the groups were comparable. Boxplots were created to visualise the distribution of scores within each condition, further confirming the similarity across groups (see Figure \ref{fig:rq0}). Thus, any differences observed in subsequent analyses can be attributed to the interventions rather than pre-existing disparities among participants.

\subsection{RQ1 -- Changes in Effectiveness and Efficiency between Phases}

\begin{figure}
    \centering
    \includegraphics[width=1\linewidth]{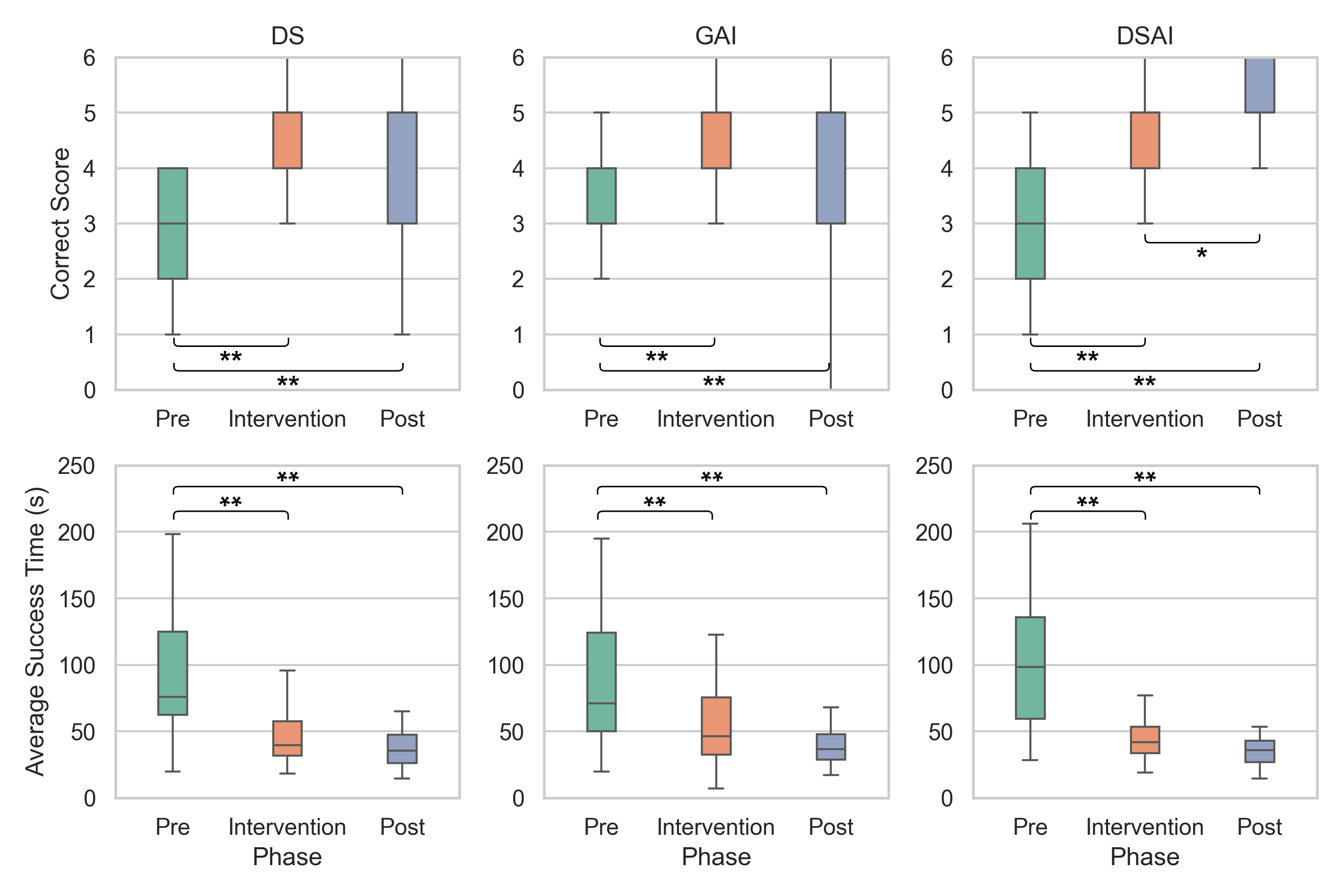}
    \caption{Changes in correct scores and success time across phases for different intervention conditions. * \textit{p} < 0.05 ** \textit{p} < 0.001}
    \label{fig:rq1}
\end{figure}

Figure \ref{fig:rq1} illustrates the within-subject changes across the three phases for each intervention condition. For the DS condition (N=36), the Friedman test revealed a significant difference in correct scores across the three phases, \(\chi^2(2) = 30.05, p < .001, W = 0.35\) (medium effect \citep{pereira2015overview}). Post-hoc Wilcoxon signed-rank tests indicated that the correct scores significantly improved from the pre-intervention (Median = 3.0, IQR = 2.0) to the intervention phase (Median = 5.0, IQR = 1.0), \(W = 46, p < .001, r = 0.91\) (large effect \citep{kerby2014simple}), and from the pre-intervention to the post-intervention phase (Median = 5.0, IQR = 2.0), \(W = 52, p < .001, r = 0.88\) (large effect). However, the change from the intervention phase to the post-intervention phase was not significant. In terms of success time for the DS condition, the Friedman test also showed significant differences, \(\chi^2(2) = 33.17, p < .001, W = 0.46\) (medium effect). Wilcoxon signed-rank tests revealed that the success time significantly decreased from the pre-intervention (Median = 75.72 seconds, IQR = 62.53 seconds) to the intervention phase (Median = 39.61 seconds, IQR = 25.96 seconds), \(W = 76, p < .001, r = 0.89\) (large effect), and from the pre-intervention to the post-intervention phase (Median = 35.58 seconds, IQR = 21.27 seconds), \(W = 22, p < .001, r = 0.97\) (large effect). The change from the intervention to the post-intervention phase was not significant. These results suggest that the DS intervention significantly improved the participants' ability to extract correct insights from data visualisations and reduced the time taken to do so. The improvements from the pre-intervention to the intervention phase and from the pre-intervention to the post-intervention phase were significant, indicating the effectiveness of the DS condition. However, there was no significant difference between the intervention and the post-intervention phase, suggesting that the gains in performance were maintained post-intervention.

For the GAI condition (N=41), the Friedman test indicated significant differences in correct scores, \(\chi^2(2) = 29.56, p < .001, W = 0.29\) (small effect). Post-hoc tests showed significant improvements from the pre-intervention (Median = 3.0, IQR = 1.0) to the intervention phase (Median = 4.0, IQR = 1.0), \(W = 71, p < .001, r = 0.87\) (large effect), and from the pre-intervention to the post-intervention phase (Median = 5.0, IQR = 2.0), \(W = 104, p < .001, r = 0.81\) (large effect). The change from the intervention to the post-intervention phase was not significant. The success time for the GAI condition also showed significant differences, \(\chi^2(2) = 40.59, p < .001, W = 0.49\) (medium effect). Wilcoxon signed-rank tests indicated significant reductions from the pre-intervention (Median = 72.12 seconds, IQR = 77.27 seconds) to the intervention phase (Median = 46.34 seconds, IQR = 42.88 seconds), \(W = 57, p < .001, r = 0.93\) (large effect), and from the pre-intervention to the post-intervention phase (Median = 36.8 seconds, IQR = 19.74 seconds), \(W = 86, p < .001, r = 0.90\) (large effect). The change from the intervention to the post-intervention phase was not significant. These findings indicate that the GAI intervention was effective in enhancing participants' correct scores and reducing their success times from the baseline to both the intervention and the post-intervention phase. However, the lack of significant differences between the intervention and the post-intervention phase suggests that the improvements were sustained post-intervention, similar to the DS condition.

For the DSAI condition (N=40), the Friedman test revealed significant differences in correct scores, \(\chi^2(2) = 46.48, p < .001, W = 0.50\) (large effect). Post-hoc Wilcoxon signed-rank tests indicated significant improvements from the pre-intervention (Median = 3.0, IQR = 2.0) to the intervention phase (Median = 5.0, IQR = 1.0), \(W = 32, p < .001, r = 0.94\) (large effect), and from the pre-intervention to the post-intervention phase (Median = 6.0, IQR = 1.0), \(W = 19, p < .001, r = 0.97\) (large effect). Additionally, there was a significant improvement from the intervention to the post-intervention phase, \(W = 209, p = .019, r = 0.52\) (large effect). Significant differences were also found in the success time for the DSAI condition, \(\chi^2(2) = 43.65, p < .001, W = 0.55\) (large effect). Wilcoxon signed-rank tests indicated significant reductions from the pre-intervention (Median = 98.28 seconds, IQR = 76.31 seconds) to the intervention phase (Median = 41.94 seconds, IQR = 19.77 seconds), \(W = 20, p < .001, r = 0.98\) (large effect), and from the pre-intervention to the post-intervention phase (Median = 35.75 seconds, IQR = 16.06 seconds), \(W = 42, p < .001, r = 0.95\) (large effect). The change from the intervention to the post-intervention phase was not significant. These results demonstrate significant improvements in both correct scores and success times from the baseline to the intervention and the post-intervention phase. Unlike the other conditions, the DSAI intervention also showed significant improvement from the intervention to the post-intervention phase, indicating a continued positive improvement post-intervention.

\subsection{RQ2 -- Changes in Effectiveness and Efficiency between Interventions}

\begin{figure}
    \centering
    \includegraphics[width=1\linewidth]{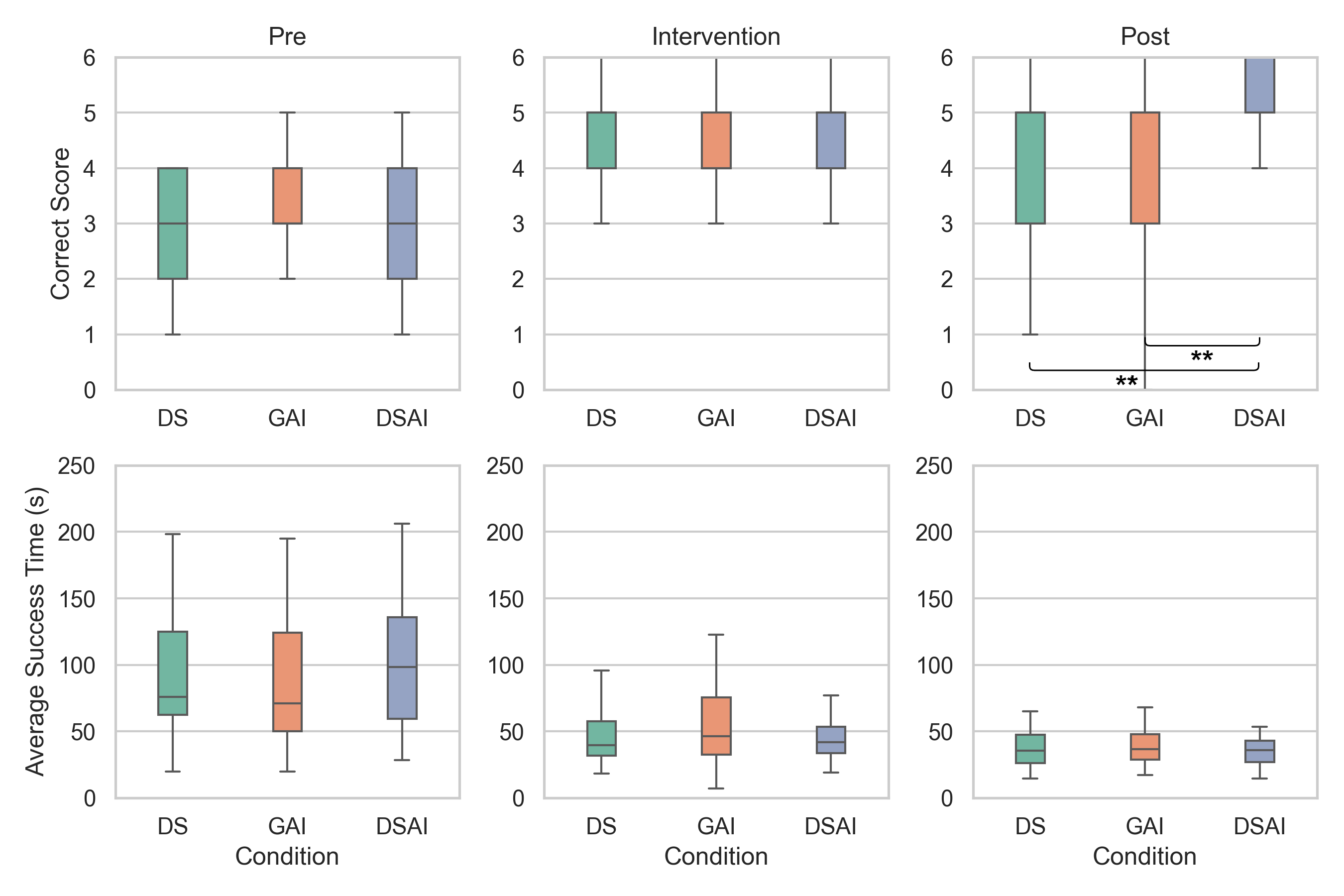}
    \caption{Comparison of correct scores and success times across intervention conditions for each phase. * \textit{p} < 0.05 ** \textit{p} < 0.01}
    \label{fig:rq2}
\end{figure}

Figure \ref{fig:rq2} illustrates the between-subject changes across the three conditions for each intervention phase. The Kruskal-Wallis test did not reveal significant differences between the three conditions for the correct scores in both the pre-intervention and the intervention phase. However, for the post-intervention correct scores, the Kruskal-Wallis test revealed significant differences, \(H(2) = 10.62, n = 117, p = .005, \eta^2 = 0.08\) (medium effect). Pairwise comparisons using the Mann-Whitney U test indicated significant differences between the DSAI and DS conditions (\(U = 967.5, p = .007, r = 0.30\); medium effect), and between the DSAI and GAI conditions (\(U = 1119.0, p = .003, r = 0.31\); medium effect). Specifically, the median correct score for DSAI was 6.0 (IQR = 1.0), which was significantly higher than the median scores for DS (Median = 5.0, IQR = 2.0) and GAI (Median = 5.0, IQR = 2.0). For the success times, the Kruskal-Wallis tests revealed no significant differences between the intervention conditions across all three phases. These between-subject results suggest that while all three interventions (DS, GAI, and DSAI) were effective in improving correct scores and reducing success times within subjects (from RQ1 findings), the DSAI condition was particularly effective in achieving higher correct scores in the post-intervention phase compared to the DS and GAI conditions. This indicates that the DSAI intervention may be more effective in sustaining performance improvements post-intervention. There were no significant differences in success times between the conditions, suggesting that all interventions were similarly effective in reducing the time taken to extract insights from the visualisations.

\subsection{RQ3 -- Visualisation literacy}

\subsubsection{Intervention score and time.} The OLS regression revealed no significant main effects or interaction effects for visualisation literacy on the intervention score for all three conditions. Regarding the success time, the OLS regression also did not reveal any significant main or interaction effects for visualisation literacy across the DS, GAI, or DSAI conditions. 

\subsubsection{Post-intervention score and time.} The regression revealed a significant main effect for visualisation literacy (\(\beta = 0.41, SE = 0.17, t = 2.33, p = 0.022\)) and the DSAI condition (\(\beta = 0.55, SE = 0.25, t = 2.52, p = 0.013\)) on the post-intervention score. This indicates that an increase of one standard deviation in visualisation literacy results in a 0.405 standard deviation increase in the post-intervention score across all three conditions. Additionally, being in the DSAI condition results in a 0.546 standard deviation increase in the post-intervention score compared to the DS and GAI conditions while holding visualisation literacy constant. The model explained 27.7\% of the variance in post-intervention scores (\(F(5, 93) = 7.11, p < 0.001\)). Whereas, the OLS regression for post-intervention time revealed no significant main or interaction effects for visualisation literacy on the post-intervention time for the DS and GAI conditions. However, the model is significant but merely explained 11.8\% of the variance in post-intervention scores ((\(F(5, 93) = 2.35, p = 0.047\))).

\section{Discussion}
\subsection{Summary of Findings and Research Questions}

As VLA become more integrated into educational settings \citep{vieira2018visual}, it is imperative to develop practical and scalable strategies to enhance the comprehension of these complex visualisations. Our study specifically addresses the challenge many students and educators face due to a lack of data visualisation literacy \citep{donohoe2020data, pozdniakov2023teachers}. By evaluating the impact of three intervention approaches, data storytelling, passive GenAI agents, and proactive GenAI agents, our research provides empirical insights into their effectiveness and efficiency in improving students' understanding of VLA. These findings align with the growing need to support educational stakeholders in navigating intricate VLA tools without overwhelming cognitive load, thereby advancing instructional strategies and educational experiences \citep{verbert2020learning}.

Regarding the first research question, our findings show that all three interventions significantly improved both the effectiveness (measured by correct scores) and efficiency (measured by average success time) of extracting insights from VLA. These improvements persisted post-intervention for data storytelling and passive GenAI agents, while proactive GenAI agents showed continued enhancement even after their removal. This provides empirical to support the use of GenAI agents as viable alternatives to data storytelling for facilitating accurate and efficient comprehension, echoing recent recommendations to harness advanced AI systems for understanding data visualisations \citep{li2024we, fernandez2024data}. Furthermore, the ongoing improvements after intervention removal suggest these methods not only provide temporary assistance but also contribute to the learning of vital visualisation comprehension skills, extending their educational benefits \citep{asamoah2022improving, bahtaji2020improving}.

In terms of the second research question, we found that the proactive GenAI agent infused with scaffolding techniques significantly outperformed both data stories and the passive GenAI agent in the effectiveness of extracting insights from data visualisations after the intervention was removed. This finding supports the effectiveness of scaffolding techniques for mastering a subject \citep{van2010scaffolding, kim2018effectiveness}, which in this study pertains to students' ability to comprehend VLA. As these techniques are also frequently used in the creation of data stories \citep{shao2024data, fernandez2024data}, the proactive agent can be viewed as a combination of scaffolding (e.g., data storytelling) and the passive GenAI agent. It actively conveys data narratives by prompting students with targeted questions instead of passively responding to students' prompts. In this sense, the current finding also provides empirical evidence to support the combined effects of integrating scaffolding with GenAI technologies to achieve better comprehension and learning skills in the interpretation of VLA. 

For the third research question, the absence of interactional effects between intervention conditions and visualisation literacy on participants' effectiveness in understanding VLA is consistent with \citet{shao2024data}'s findings. This indicates that all three interventions are equally beneficial for students, regardless of their level of visualisation literacy. Contrary to the common belief supported by previous studies \citep{Zhang2022framework, pozdniakov2023teachers}, our results did not statistically confirm that any supportive methods, including data storytelling, are more advantageous for individuals with lower visualisation literacy. Furthermore, the significant results demonstrating the proactive GenAI agent's ability to enhance participants' effectiveness in comprehending VLA after the intervention was removed, while accounting for visualisation literacy, further validate its benefits for students across all levels of visualisation literacy.

\subsection{Implications for Educational Research}

This study has significant implications for the fields of education and learning analytics. Our findings suggest that leveraging advanced GenAI technologies can provide effective alternatives or complements to conventional scaffolding methods like data storytelling, thereby enhancing students' understanding of complex visual analytics in educational settings \citep{li2024we, yan2024vizchat}. Specifically, the proactive GenAI agent's superior post-intervention performance highlights the potential of scaffolding techniques for cultivating essential comprehension skills in VLA \citep{gibbons2002scaffolding, kim2018effectiveness}. Researchers are encouraged to explore the development of these GenAI agents further to offer adaptive, context-aware support that evolves alongside students' growing analytical skills \citep{asamoah2022improving, bahtaji2020improving}.

The positive effects observed with GenAI agents do not diminish the value of conventional scaffolding methods like data storytelling. Instead, they reinforce the effectiveness of the principal design elements of data stories \citep{Segel2010, Zdanovic2022}, which align with established educational scaffolding techniques \citep{van2010scaffolding, kim2018effectiveness}. However, the current study focuses on improved effectiveness and efficiency in comprehending insights from visual analytics and does not necessarily equate to enhanced memory retention, learner engagement, or emotional reaction to VLA \citep{vieira2018visual, noroozi2019multimodal}. Thus, our study advocates for a broader research initiative into human-AI collaboration in designing educational tools that both support users in interpreting VLA and foster the development of comprehension skills. This has become increasingly important across various educational contexts \citep{vieira2018visual, verbert2020learning}.

The lack of significant interactional effects between visualisation literacy and intervention conditions aligns with \citet{shao2024data}'s findings, suggesting that visualisation literacy may not be as pivotal as previously thought in determining the effectiveness of supportive interventions \citep{pozdniakov2023teachers}. Nevertheless, it remains a crucial factor once external support is removed, paving the way for more nuanced investigations into how visualisation literacy interacts with supportive tools to aid comprehension. Further research should also consider other individual differences, like cognitive load or domain-specific expertise, that might impact the effectiveness of these educational interventions.

\subsection{Implications for Educational Practice}

This study offers actionable insights for improving data visualisation comprehension in educational environments. Integrating GenAI agents and data storytelling into educational tools can significantly aid students and educators in extracting and understanding insights from complex visual analytics \citep{yan2024vizchat, ma2023demonstration}. The improvements in effectiveness and efficiency of these methods can facilitate more informed decision-making and understanding in educational contexts, ensuring timely and accurate comprehension of insights communicated via VLA \citep{vieira2018visual}. The sustained post-intervention improvements highlight their role in educational settings, particularly for students, who often struggle with low visualisation literacy \citep{donohoe2020data}.

As these methods are effective regardless of visualisation literacy levels, educators should consider them as educational tools that go beyond temporary aids. Training programs and workshops could incorporate these technologies to deepen students' understanding of visualisation principles. Notably, the proactive GenAI agent’s ability to sustain improvements post-intervention suggests it offers unique advantages for enhancing both learning and performance \citep{soderstrom2015learning}. It is particularly useful for guiding students through complex VLA, especially when integrating insights from multiple sources is required \citep{noroozi2019multimodal, martinez2020data}. Such learning processes are valuable for educational programmes across various domains, such as science, business, and social studies, where interpreting visual data is critical for learning and knowledge development \citep{vieira2018visual, sahin2021visualizations}.

However, integrating GenAI technologies into educational settings must be approached with caution, particularly concerning data privacy and security \citep{yan2024practical, zohny2023ethics}. Establishing knowledge databases to store contextual information for GenAI systems may raise privacy concerns, necessitating compliance with regulatory standards to ensure ethical application.

\subsection{Limitations and Future Directions}

While this study provides valuable insights, certain limitations should be addressed in future research. Firstly, the controlled experimental environment may not fully mirror the complexities encountered in real-world educational settings involving VLA. To gain a comprehensive understanding, future research should conduct field experiments or longitudinal studies to assess the effectiveness of these interventions in authentic educational contexts. Secondly, the participant pool was predominantly from North America and Europe due to recruitment limitations. For future studies, incorporating a more diverse participant group is essential to ensure broader applicability of the findings. Thirdly, although the study demonstrated the promise of proactive GenAI agents, it did not thoroughly investigate the specific design features and interaction mechanisms contributing to their effectiveness. Future research should delve into the design and implementation of these agents, focusing on enhancing scaffolding techniques and feedback mechanisms to optimise students' comprehension of VLA. Moreover, although the findings should be generalisable to other VLA applications in different educational contexts, given the involvement of participants with limited contextual knowledge in this study, future research should evaluate the interventions' effectiveness in promoting higher levels of Bloom's Taxonomy beyond knowledge (Level 1) and comprehension (Level 2). Additionally, we did not analyse students' analytical writing content due to potential subjectivity and its limited relevance to the study's primary focus. Future work will utilise automated essay grading and learning analytics to gain deeper insights into students' interactions with supportive methods and their impacts on the writing process \citep{ramesh2022automated, kovanovic2017content}.
\section{Conclusion}

This study advances our understanding of how data storytelling, passive GenAI agents, and proactive GenAI agents enhance students' comprehension of VLA. Addressing our research questions, we found that all three methods significantly improved both the effectiveness and efficiency of extracting insights from VLA, with proactive GenAI agents demonstrating unique advantages. Unlike previous methods, proactive agents not only maintained comprehension improvement but also further enhanced it post-intervention, showcasing their ability to nurture lasting comprehension skills. These findings contribute to existing pedagogical theories by confirming that the integration of advanced GenAI technologies with traditional scaffolding techniques, such as data storytelling, can substantially benefit educational stakeholders. Our results offer empirical evidence that advances theories concerning the pedagogical use of GenAI agents, illustrating that proactive GenAI agents can be designed to respond dynamically to and guide students throughout their learning process, thus facilitating deeper understanding and genuine learning.










\section*{Acknowledgement}
This study was in part supported by grants from the Australian Research Council (grant agreement numbers DP220101209 and DP240100069 to D.G.). L.Y.'s work is fully funded by the Digital Health CRC (Cooperative Research Centre). D.G.'s work was, in part, supported by the DHCRC and Defense Advanced Research Projects Agency (DARPA) through the Knowledge Management at Speed and Scale (KMASS) program (HR0011-22-2-0047). The DHCRC is established and supported under the Australian Government's Cooperative Research Centres Program. The U.S. Government is authorised to reproduce and distribute reprints for Governmental purposes notwithstanding any copyright notation thereon. The views and conclusions contained herein are those of the authors and should not be interpreted as necessarily representing the official policies or endorsements, either expressed or implied, of DARPA or the U.S. Government.

\bibliographystyle{cas-model2-names}

\bibliography{0_reference}


\appendix
\section{Scaffolding Prompts and Questions}
\label{appendix:scaffold}
\subsection{Scaffolding Prompts}
\textbf{Scaffolding Approach.} Guide participants through understanding each visualisation step-by-step. For each visualisation, start by providing a one-sentence description and ask only one question at a time. Avoid asking repeated questions. Give feedback on participants' responses. Once all questions for one visualisation are covered, direct participants to the next visualisation by asking them to click the right arrow. Only ask questions related to the visualisation that was sent to you. Ensure you cover all the \textbf{Scaffolding Questions} for each visualisation, including 1 for the Bar Chart, 2 for the Communication Network, and 3 for the Ward Map. Use a friendly and conversational tone. If participants get something wrong, provide a hint and ask them to rethink and re-answer. 

\subsection{Visualisation Descriptions and Scaffolding Questions}
\noindent\textbf{Bar Chart}
\begin{itemize}
    \item \textbf{Description:} It shows the proportion of time the two nurses spent on different tasks.
    \item \textbf{Scaffolding Question:}
    \begin{enumerate}
        \item What do you notice about Amy's proportion of time spent on tasks, and what does this indicate about her prioritisation?
    \end{enumerate}
\end{itemize}

\noindent\textbf{Communication Network}
\begin{itemize}
    \item \textbf{Description:} It shows verbal communication interactions among each student and their communication with the doctor, relative, and patient manikins. Arrows indicate the direction of communication, and edge thickness indicates the duration of communication.
    \item \textbf{Scaffolding Questions:}
    \begin{enumerate}
        \item Who are the main communicators in the network, and what does the thickness of the arrows tell us?
        \item How often did the nurses communicate with each other compared to the doctor, patient, and relative?
    \end{enumerate}
\end{itemize}

\noindent\textbf{Ward Map}
\begin{itemize}
    \item \textbf{Description:} It shows the verbal and spatial distribution of each student during a simulation session. Saturated colours indicate frequent verbal communication, while the hexagons' locations show spatial distributions. The peak heart rate of each student is also displayed on the Ward Map, represented by a heart shape.
    \item \textbf{Scaffolding Questions:}
    \begin{enumerate}
        \item Which areas of the ward did the nurses spend the most time in, and how does that relate to their task prioritisation?
        \item How can the colour intensity give us insights into verbal communication patterns?
        \item What does the peak heart rate tell us about the stress levels of the nurses in different areas of the ward?
    \end{enumerate}
\end{itemize}

\section{Evaluation Questions}
\label{appendix:evaluation}

\begin{table*}[htbp]
\centering
\caption{Evaluation Questions}
{
\renewcommand{\arraystretch}{2}
\begin{tabular}{llp{11.5cm}}
\hline
\textbf{Visualisation} & \textbf{Bloom's Level} & \textbf{Evaluation Question} \\ \hline
Bar Chart & Knowledge & \textbf{Q:} Which behaviour did the two nurses spend the \textit{least} time on? \newline 
-- Working together on other tasks \newline 
-- Working individually on other tasks \newline 
-- Working together on tasks for Amy \newline 
-- Working individually on tasks for Amy \\ \hline
Bar Chart & Comprehension & \textbf{Q:} How did the nurses spend their time working on tasks for Amy compared to other tasks? \newline 
-- Two nurses spent more time working individually on tasks for Amy than on other tasks. \newline 
-- Two nurses spent more time working together on other tasks than on tasks for Amy. \newline 
-- Two nurses spent the same amount of time working individually on other tasks and on tasks for Amy. \newline 
-- Two nurses spent the same amount of time working on other tasks and on tasks for Amy. \\ \hline
Social Network & Knowledge & \textbf{Q:} Who did the Primary Nurse 2 communicate with? \newline 
-- Secondary Nurse 1, Patient, and Doctor \newline 
-- Secondary Nurse 1, Patient, and Relative \newline 
-- Primary Nurse 1, Doctor, and Relative \newline 
-- Primary Nurse 1, Patient, and Relative \\ \hline
Social Network & Comprehension & \textbf{Q:} Which of the following statements is correct? \newline 
-- Primary Nurse 2 communicated more with the Relative than Primary Nurse 1 \newline 
-- Primary Nurse 2 communicated more with the Patient than Primary Nurse 1 \newline 
-- Primary Nurse 1 communicated more with the Relative than Primary Nurse 2 \newline 
-- Primary Nurse 1 communicated more with the Patient than Primary Nurse 2 \\ \hline
Ward Map & Knowledge & \textbf{Q:} Which of the following statements best describes the behaviours of the two nurses? \newline 
-- The two nurses spent a significant amount of time but did not talk much around Bed 4. Primary Nurse 1's heart rate peaked at 131 around Bed 4. \newline 
-- The two nurses spent a significant amount of time but did not talk much around Bed 3. Primary Nurse 2's heart rate peaked at 95 around Bed 4. \newline 
-- The two nurses spent a significant amount of time and talked a lot around Bed 4. Primary Nurse 1's heart rate peaked at 131 around Bed 4. \newline 
-- The two nurses spent a significant amount of time and talked a lot around Bed 3. Primary Nurse 2's heart rate peaked at 95 around Bed 4. \\ \hline
Ward Map & Comprehension & \textbf{Q:} Which of the following statements is correct? \newline 
-- Primary Nurse 1's spatial and verbal activities were more around Bed 1 than Primary Nurse 2's. Primary Nurse 1 had a higher increase in heart rate than Primary Nurse 2. \newline 
-- Primary Nurse 1's spatial and verbal activities were more around Bed 2 than Primary Nurse 2's. Primary Nurse 1 had a higher increase in heart rate than Primary Nurse 2. \newline 
-- Primary Nurse 2's spatial and verbal activities were more around Bed 1 than Primary Nurse 1's. Primary Nurse 2 had a higher increase in heart rate than Primary Nurse 1. \newline 
-- Primary Nurse 2's spatial and verbal activities were more around Bed 2 than Primary Nurse 1's. Primary Nurse 2 had a higher increase in heart rate than Primary Nurse 1. \\ \hline
\end{tabular}
}
\end{table*}
\end{document}